\documentstyle[emulateapj]{article}
\begin{document}
\submitted{Revised version November 11, 1999}
\title{Prompt Iron Enrichment, Two $r$-Process Components, and
Abundances in Very Metal-Poor Stars}
\author{G. J. Wasserburg\altaffilmark{1} and Y.-Z. Qian\altaffilmark{2}}
\altaffiltext{1}{The Lunatic Asylum,
Division of Geological and Planetary Sciences, California
Institute of Technology, Pasadena, CA 91125.}
\altaffiltext{2}{School of Physics and Astronomy, University of
Minnesota, Minneapolis, MN 55455; qian@physics.umn.edu.}

\begin{abstract}
We present a model to explain the wide range of abundances for
heavy $r$-process elements (mass number $A>130$)
at low [Fe/H]. This model requires
rapid star formation and/or an initial population of
supermassive stars in the earliest condensed clots of matter
to provide a prompt or initial Fe inventory. Subsequent Fe and $r$-process
enrichment was provided by two types of supernovae: one producing
heavy $r$-elements with no Fe on a rather short timescale and
the other producing light $r$-elements
($A\leq 130$) with Fe on a much longer
timescale.
\end{abstract}

\section{Introduction}

We present a phenomenological model for the abundances
of Fe and heavy (mass number $A>130$)
and light ($A\leq 130$) $r$-process elements ($r$-elements)
in very metal-poor stars.
These stars formed early in Galactic history when only a small
number of massive stars had evolved to become Type II supernovae 
and added heavy elements to the interstellar medium (ISM). Observations
(e.g., McWilliam et al. 1995; McWilliam 1998)
show that there is a lack of correlation between
the ``metallicity'' measured by [Fe/H] and 
abundances of $r$-elements above $A\sim 135$
(e.g., Ba and Eu)
for stars with $-3.1\lesssim{\rm [Fe/H]}\lesssim -2.5$. This 
disagrees with the view that Fe and $r$-elements 
are always coproduced by supernovae. 
Our approach to account for
this result is based
on a two-component $r$-process model (Wasserburg, Busso, \& Gallino
1996; Qian, Vogel, \& Wasserburg 1998;
Qian \& Wasserburg 1999, hereafter QW99) which attributes
heavy $r$-nuclei to high frequency supernovae
(H events) and light $r$-nuclei to 
less frequent ones (L events).
We propose that prompt production of Fe with minor coproduction of
heavy $r$-elements
in the earliest era of the Galaxy first enriched the ISM
up to [Fe/H]~$\sim -3$. Subsequent production
of $A>130$ nuclei with negligible coproduction of Fe
by H events then resulted in a
wide range of abundances for heavy $r$-elements at
$-3.1\lesssim{\rm [Fe/H]}\lesssim -2.5$.
We associate further Fe enrichment 
of the ISM for [Fe/H]~$> -3$ with L events.
Addition of H and L events over a sufficiently long timescale then led to 
a correlation between [Fe/H] and 
abundances of heavy $r$-elements at [Fe/H]~$\gtrsim -2.5$.

Mathews, Bazan, \& Cowan (1992) discussed using Galactic chemical
evolution to constrain the site of the $r$-process.
A number of recent studies (Ishimaru \&
Wanajo 1999; Tsujimoto, Shigeyama, \& Yoshii 1999;
McWilliam \& Searle 1999; see also Raiteri et al. 1999) focused
on the relationships between [Fe/H] and abundances of heavy
$r$-elements in the early Galaxy. 
A common consensus is that
chemical enrichment of the ISM at very early times was grossly
inhomogeneous and diverse yields of individual supernova events 
had dramatic effects on abundances in very metal-poor stars. Our
present work differs from previous studies in that the diversity in
$r$-process
production by supernovae is introduced through the two-component
$r$-process model based on solar system data independent of the
stellar observations. Further, we propose
a prompt mechanism for Fe production considering
special conditions of the early Galaxy. 
In \S2 we describe the framework of two $r$-process components. In \S3 we
use this together with
a postulated prompt Fe source to explain the observational results
on abundances (especially for Ba and Eu) in the early Galaxy. 
We discuss prompt
Fe production and give conclusions in \S4.

\section{Two $r$-process components}

Observations by Sneden et al. (1996, 1998)
demonstrated that abundances of $r$-elements in
the Pt peak ($A\sim 195$) and down to Ba ($A\sim 135$) in
CS~22892--052 ([Fe/H]~$=-3.1$), HD~115444 ([Fe/H]~$=-2.77$), and
HD~126238 ([Fe/H]~$=-1.67$)
are in remarkable accord with
the solar system $r$-process abundance pattern 
(the solar $r$-pattern). Assuming that a single $r$-pattern
extends from Ba to the actinides 
above the Pt peak, Cowan et al.
(1997, 1999) discussed using the 
Th/Eu ratio as a Galactic chronometer. However,
the discovery of $^{182}$Hf (lifetime
$\bar\tau_{182} = 1.30 \times
10^7$~yr) in meteorites
with ($^{182}$Hf/$^{180}$Hf)$_{\rm SSF}=2.4\times 10^{-4}$
(Harper \& Jacobsen 1996; Lee \& Halliday 1995, 1997)
at the time of solar system formation (SSF)
provided a new twist to our understanding of the $r$-process.
Although both $^{182}$Hf and $^{129}$I
($\bar\tau_{129}=2.27\times 10^7$~yr) are produced essentially only
by the $r$-process, the $^{182}$Hf data and abundance ratio
($^{129}$I/$^{127}$I)$_{\rm SSF}=10^{-4}$ (Reynolds 1960;
see also Brazzle et al. 1999)
cannot be explained by a single type of $r$-process events.
Based on this, Wasserburg et al.
(1996) concluded that there had to be at least two
distinct types of $r$-process events: one (H) occurring 
on a timescale $\Delta_{\rm H} \sim 10^7$~yr,
commensurate with that for replenishment of a typical molecular cloud
with fresh supernova debris,
and the other (L) occurring on a much longer timescale
$\Delta_{\rm L} \sim 10^8$~yr. 
They further pointed out that 
relative to the solar $r$-pattern, there should be
frequent abundance excesses of $r$-elements
in the Pt peak over those in the $A\sim 130$ peak below Ba
at low metallicities. 
Qian et al. (1998)
showed that in a two-component model
to account for
the solar $r$-pattern, the total mass yield of an L event
must be $\sim 10$ times that of an H event. They
also showed that it is not readily
possible to produce the $A\sim 130$ peak without substantially
populating the region beyond this peak. They further speculated that
H events might be associated with production of black
holes and L events with production of neutron stars.

A preliminary report on the observed Ag ($A\sim 107$)
abundance in CS 22892--052
by Cowan \& Sneden (1999)
appeared to support the meteoritic prediction that more than one type
of $r$-process events may be required. Based on this,
Qian \& Wasserburg
(QW99) sought to establish a quantitative basis for
predicting the yields of $r$-elements in H and L events. 
Their model uses the following assumptions:
(1) each H or L event has fixed $r$-process yields;
(2) the products of individual H and L events are mixed with a
``standard'' mass of hydrogen in the ISM; and
(3) the solar system inventory of all $r$-nuclei
(both stable and radioactive) are the result of production by H
and L events over a time
$T_{\rm UP}\approx 10^{10}$~yr preceding SSF.
The timescales for occurrence of H and L events
in a volume corresponding
to the standard mixing mass of hydrogen are $\Delta_{\rm H}$ and
$\Delta_{\rm L}$.
Under the above assumptions, at the time of SSF, the number of nuclei
for a stable nuclide ${\cal{S}}$
(e.g., $^{127}$I or $^{182}$W) mixed with a standard mass
of hydrogen is
\begin{equation}
N_{\cal{S}}(t_{\rm SSF})= Y_{\cal{S}}^{\rm H}
{T_{\rm UP}\over\Delta_{\rm H}} + Y_{\cal{S}}^{\rm L}
{T_{\rm UP}\over\Delta_{\rm L}},
\label{ns}
\end{equation}
while that for a radioactive nuclide ${\cal{R}}$
with $\bar\tau_{\cal{R}}\ll T_{\rm UP}$
(e.g., $^{129}$I or $^{182}$Hf) is
\begin{equation}
N_{\cal{R}}(t_{\rm SSF})\approx {Y_{\cal{R}}^{\rm H}
\exp(-\vartheta\Delta_{\rm H}/\bar\tau_{\cal{R}})\over
1-\exp(-\Delta_{\rm H}/\bar\tau_{\cal{R}})} + {Y_{\cal{R}}^{\rm L}
\over\exp(\Delta_{\rm L}/\bar\tau_{\cal{R}}) -1}.
\label{nr}
\end{equation}
In equations (\ref{ns}) and (\ref{nr}) $Y^{\rm H}$
or $Y^{\rm L}$ denotes the yield in an H or L event. The
yield for a radioactive species is taken to be about the same 
as that for the corresponding stable nuclide.
In equation (\ref{nr}) the interval between SSF and the last L event
is assumed to be $\Delta_{\rm L}$, while that between SSF and the
last H event is specified by $\vartheta$
$(0\leq\vartheta\leq 1)$. From the meteoritic data on
$^{129}$I and $^{182}$Hf, it was found that
$\Delta_{\rm H}\leq 1.85\times 10^7$~yr and
$\Delta_{\rm L}\geq 1.06\times 10^8$~yr for $\vartheta=1$ (scenario A).
Further, the yields for $^{127}$I and
$^{182}$W in an L event relative to those in an H event,
$Y_{127}^{\rm L}/Y_{127}^{\rm H}$ and $Y_{182}^{\rm L}/Y_{182}^{\rm H}$,
are rather strongly constrained for
all cases bracketed by scenarios A
and B ($\vartheta=0$). Given $\vartheta$,
$\Delta_{\rm H}$, and $\Delta_{\rm L}$, the yields for
$^{127}$I and $^{182}$W in an H or L event can be obtained from
the relevant solar system data through equations (\ref{ns}) and
(\ref{nr}).

To generalize the results to other 
$r$-nuclei, Qian \& Wasserburg (QW99) assumed that the yield
template for $A>130$ nuclei associated with $^{182}$W is the same
as the corresponding solar $r$-pattern for both H and L events
based on the observations by Sneden
et al. (1996, 1998). The yields for $A\leq 130$
nuclei associated with $^{127}$I in an H or L event were also chosen
to follow the corresponding solar $r$-pattern. Consequently,
the yield for a stable
nucleus ${\cal{S}}$ with mass number $A$
in an L event relative to that in an H event is
$Y_{\cal{S}}^{\rm L}/Y_{\cal{S}}^{\rm H}=Y_{127}^{\rm L}/Y_{127}^{\rm H}$ 
for $A\leq 130$ or
$Y_{\cal{S}}^{\rm L}/Y_{\cal{S}}^{\rm H}=Y_{182}^{\rm L}/Y_{182}^{\rm H}$
for $A>130$.
The fraction of $\cal{S}$ nuclei contributed by
H or L events to the corresponding solar $r$-process abundance
are $F_r^{\rm H}({\cal{S}})
=1/[1+(Y_{\cal{S}}^{\rm L}/Y_{\cal{S}}^{\rm H})
(\Delta_{\rm H}/\Delta_{\rm L})]$
or $F_r^{\rm L}({\cal{S}})=1-F_r^{\rm H}({\cal{S}})$.
From equation (\ref{ns}) 
the abundance of stable ${\cal{S}}$ nuclei
resulting from a single H or L event contaminating a standard
mass of hydrogen in the ISM is
$\log\epsilon_{\rm H}({\cal{S}}) = \log\epsilon_{\odot,r}({\cal{S}})
+\log F_r^{\rm H}({\cal{S}}) - \log(T_{\rm UP}/\Delta_{\rm H})$ or
$\log\epsilon_{\rm L}({\cal{S}}) = \log\epsilon_{\odot,r}({\cal{S}})
+\log F_r^{\rm L}({\cal{S}}) - \log(T_{\rm UP}/\Delta_{\rm L})$.
Here the spectroscopic notation
$\log\epsilon({\cal{S}})\equiv\log({\cal{S}}/{\cal{H}})+12$ is used
(${\cal{S}}/{\cal{H}}$ being the number abundance ratio 
of ${\cal{S}}$ to hydrogen).
Thus, given $\vartheta$, $\Delta_H$, and $\Delta_L$,
there is a
quantitative prediction for the abundances resulting from a single
$r$-process event. For example, a single H event gives rise to
$\log\epsilon_{\rm H}({\rm Eu})\approx -3.0$ to $-2.2$ and
$\log\epsilon_{\rm H}({\rm Ba})\approx -2.1$ to $-1.3$ over
a wide range of model parameters.
The results from a mixture of multiple events can be calculated
simply by adding the number
of nuclei produced in each event and then converting it to the
corresponding $\log\epsilon$ value for the mixture (see QW99). 

\section{The Iron Conundrum}

Observational data (Gratton \& Sneden 1994;
McWilliam et al. 1995; McWilliam 1998;
Sneden et al. 1996, 1998) on Ba/Eu,
$\log\epsilon({\rm Eu})$, and [Fe/H] for low-metallicity
stars are shown in Figure 1. 
As can be seen, 
there is a wide dispersion in $\log\epsilon({\rm Eu})$
at $-3.1\lesssim{\rm [Fe/H]}\lesssim -2.5$
while Ba/Eu is
essentially constant. The $r$-process accounts for over 90\%
of the solar Eu inventory 
but only about 20\% of the solar Ba inventory
(K\"appeler, Beer, \& Wisshak 1989;
Arlandini et al. 1999). Consequently, 
the clustering of Ba/Eu around
the solar $r$-process value exhibited in Figure 1 confirms the
earlier proposal by Truran (1981) that heavy elements such as
Ba in very metal-poor stars originated from the $r$-process.
There are few Eu data at
$-4\lesssim{\rm [Fe/H]}\lesssim -3$. However, sufficient
Ba data at
these metallicities (McWilliam et al. 1995; McWilliam 1998)
are available and shown in Figure 2 (region A).
Both Ba and Eu data show a wide dispersion 
at $-3.1\lesssim{\rm [Fe/H]}\lesssim -2.5$ (region B) and suggest that
Fe and heavy $r$-nuclei are not coproduced by common
supernovae (H events).
The $\log\epsilon({\rm Ba})$ and $\log\epsilon({\rm Eu})$ values
from a single H event in our model (QW99) are indicated
by the zone marked ``1 H'' in the corresponding figure.
We expect that increases of $\log\epsilon({\rm Ba})$
and $\log\epsilon({\rm Eu})$
above this zone are dominantly the results of adding more H events
to the standard mixing mass of hydrogen. 
For example, if we take 
$\log\epsilon({\rm Eu})\approx -2.5$ for HD 122563 
([Fe/H]~$\approx -2.7$) as representative of
a single H event, then the Eu abundance in CS 22892--052
([Fe/H]~$\approx -3.1$) would correspond to $\sim 30$
H events. As the Fe abundance in CS 22892--052 is smaller than
that in HD 122563, it is evident that Fe cannot be significantly
produced by H events.
Thus the wide dispersion in Ba and Eu abundances at [Fe/H]~$\sim -3$
poses a conundrum of Fe production in the early Galaxy
and suggests that
[Fe/H] is neither related to heavy $r$-element production nor a reliable
chronometer (see QW99).

We find that the Fe conundrum
can be resolved by postulating an initial or promptly-generated
Fe inventory that existed before the occurrence of
H and L events. By ``initial'' we mean very early stages during
which an inventory of Fe was provided to the ISM from which
regular stars later formed with no other temporal connection. 
Prompt Fe production is considered
to be associated with coevolution of all stars from an initial 
gas clot with no Fe. 
In both cases the mechanism for Fe production
must have ceased at [Fe/H]~$\sim -3$.
Then non-Fe-producing H events 
and Fe-producing L
events (see below) began to occur.
The frequent occurrence of H events would result in a range of
abundances for heavy $r$-elements such as Ba and Eu
at [Fe/H]~$\sim -3$ while
the less frequent occurrence of L events would lead to
increases in Fe abundance above [Fe/H]~$\sim -3$.
A correlation between [Fe/H] and abundances of heavy $r$-elements
would then be established
when sufficient Fe was produced by L events to overwhelm the
inventory produced by the initial/prompt Fe source  
which was not related to ``typical'' supernovae.
The existence of some stars with $-4\lesssim{\rm [Fe/H]}\lesssim -3$ 
in region A of Figure 2
indicates that the initial/prompt Fe production had diverse yields and/or
was sufficiently extended
in time so that [Fe/H]$\sim -3$ represents the sum
of a number of individual events. 
The Ba abundances in region A could be attributed to
production by the initial/prompt Fe source which would be small compared
with that by a single H event.
However, these data could also be explained by a mixing
scenario with no Ba production by the initial/prompt Fe source (see \S4).
The onset of the correlation between [Fe/H] and abundances of
heavy $r$-elements
can be estimated as follows. By attributing $\sim 1/3$ of the solar Fe
inventory (Timmes, Woosley, \& Weaver 1995) to the Type II supernovae
associated with L events, we expect a single L event to result in
$\log\epsilon_{\rm L}({\rm Fe})\sim 5.0$ corresponding to
[Fe/H]~$\sim -2.5$ if we take
$\Delta_{\rm L}\sim 10^8$~yr ($\sim 100$ L events are then
responsible for the part of the solar Fe
inventory contributed by Type II supernovae). Therefore,
we expect that a correlation between [Fe/H]
and abundances of heavy $r$-elements would be established
through addition of H and L events over 
a few $10^8$ yr during which the ISM was sufficiently enriched by L
events to [Fe/H]~$\gtrsim -2.5$. Indeed, data
in Figures 1 and 2 (region C) show that such a correlation exists
at [Fe/H]~$\gtrsim -2.5$.

\section{Discussion and Conclusions}

There is a basic issue of what
plausible mechanism could account for the initial or promptly-generated
Fe inventory. The prompt Fe production in a gas clot 
must have lasted for only a narrow time interval
($\ll \Delta_{\rm H}\sim 10^7$~yr)
and then was greatly diminished. A possible mechanism would involve 
Fe production by supermassive stars
with no significant coproduction of heavy $r$-nuclei.
Let us consider a pristine gas clot of mass $M_0$ from which
two populations of stars could be made:
supermassive stars with very short
lifetimes ($\lesssim 10^6$~yr) and less massive
stars with lifetimes longer than
$\sim 10^6$~yr. 
So $M_F(t)+M_P(t)+M_g(t)=M_0$,
where $M_F(t)$, $M_P(t)$, and $M_g(t)$ represent the masses stored in
supermassive (``Fat'') stars, less massive (``Petite'')
stars, and gas at time $t$. 
Taking the birth rates of both types of stars
to be proportional to $M_g(t)$
and assuming that supermassive stars were 
born at about the same rate as they were destroyed, we have
$\dot M_F(t)\approx K_F M_g(t) - M_F(t)/\bar\tau_F\approx 0$, 
where $\bar\tau_F$ is the average lifetime of supermassive stars,
and $\dot M_P(t)=K_P M_g(t)$. Thus,
$M_F(t)\approx K_FM_0\bar\tau_F\exp(-K_P t)$ and $M_P(t)\approx
M_0[1-\exp(-K_P t)]$.
Therefore, if the birth rate for less massive stars was
sufficiently high to deplete the gas over a few $10^6$~yr,
then the population of supermassive stars would decline 
on the same timescale. This would provide some diversity in 
Fe abundances in stars formed at very early times 
but effectively truncate
further addition of Fe from supermassive stars.

Stars with masses $11\lesssim M/M_\odot\lesssim 40$ are considered to
become supernovae. Their lifetimes range from $6\times 10^6$~yr to
$2\times 10^7$~yr (see Fig. 1 of Timmes et al. 1995). Meynet et al.
(1994) gave a lifetime $\approx 3\times 10^6$~yr for a $120\,M_\odot$
star. Therefore, prompt Fe production in our model must be associated
with supermassive stars
of at least a few $100\,M_\odot$. These stars are assumed 
to produce no significant amount of 
heavy $r$-nuclei but sufficient Fe to give [Fe/H]~$\sim -3$
over a few $10^6$~yr.
As speculated above, the termination of prompt Fe production 
was caused by rapid depletion of gas through storage in less
massive stars over a few $10^6$~yr. Thus the initial star 
formation rate in a pristine gas clot of mass
$\sim 10^6\,M_{\odot}$
is required to be $\sim 1\,M_\odot$~yr$^{-1}$.
When extrapolated to the whole Galaxy,
this corresponds to an initial rate $\sim 10^5\,M_\odot$~yr$^{-1}$,
much higher than the average value of $\sim 10\,M_\odot$~yr$^{-1}$ over
Galactic history. On a longer timescale, stars with
$11\lesssim M/M_\odot\lesssim 40$
became supernovae and began to enrich the ISM.
These supernovae are of two types: high frequency H events 
and low frequency L events.
Within a standard mass of hydrogen, 
H and L events occur on timescales $\sim
10^7$~yr and $\sim 10^8$~yr, respectively.
The H events produce heavy $r$-elements but no Fe and this resulted
in a wide range of abundances for heavy $r$-elements (e.g., Ba and Eu) at 
[Fe/H]~$\sim -3$.
A correlation between [Fe/H] and abundances of heavy $r$-elements
was established later through addition of H and L events
over a few $10^8$~yr when
Fe production by L events overwhelmed the prompt Fe inventory.
The observed onset of this correlation at [Fe/H]~$\sim -2.5$ is 
consistent with the expected Fe yield of L events.
The L events also produce light $r$-elements
such as Ag. Production of these elements
by the hypothesized prompt Fe source is unknown. 
It was argued that the first stars of H-He composition
would be very massive (e.g., Truran \& Cameron 1971). 
These stars would provide elements heavier
than He to the ISM (e.g., Ezer \& Cameron 1971). 
How much Fe would be produced by these stars and how this Fe would
be mixed with the ISM
require more investigation.

It is likely that supermassive stars would blow up their
parent gas clots, thus preventing further star formation.
In this case they would provide Fe to the general ISM. 
We speculate that once [Fe/H]~$\sim -3$ was reached, 
supermassive stars could no longer
be produced and less massive stars would form instead. 
In this scenario an initial Fe inventory could be provided
without requiring a high star formation rate in 
the gas clots before the H and L events occurred. 
We note that the cut-off at [Fe/H]~$\sim -3$ may correspond 
to a condition when sufficient amount of elements 
heavier than He was provided by supermassive stars to permit 
adequate cooling of aggregating matter for less
massive ($\lesssim 40\,M_{\odot}$) stars to form. A supermassive 
star of a few $100\,M_{\odot}$ must produce a few $M_\odot$ of Fe
in order to give [Fe/H]~$\sim -3$ to a clot
of $\sim 10^6\,M_{\odot}$.

Region A of Figure 2 deserves special attention. We are faced with
Ba abundances
below the production by a single H event at
$-4\lesssim [{\rm Fe/H}]\lesssim -3$. 
These Ba abundances could be attributed to minor heavy $r$-element
production by
supermassive stars. However, supernova explosions could drive gas
outflows from a clot. The enriched gas could then mix with 
the pristine
gas in other clots. In this way Ba abundances below the production
of a single H event for [Fe/H]~$<-3$ could be obtained.
For example, the abundances in the star at the left corner of region A 
could be explained by a mixture of gas outflow after an H event
with the gas in a pristine clot (mixing ratio $\sim$~1:10). 
We have avoided the complexities of mixing and exchange in our previous
discussion based on a standard mixing mass. Subsequent models must address
these issues.

The chronometric interpretation of [Fe/H] is complex. Condensation of
matter to form stars during the early evolution of the Galaxy is expected
to have been greatly extended in space and time ($\sim
10^8$--$10^9$~yr). This means that clots of baryonic matter formed
within the Galaxy at widely
disparate times. The model proposed here only requires the following
sequence of events to occur within a clot: an initial Fe inventory
from or prompt Fe enrichment by
supermassive stars, enhancement in heavy $r$-elements
by H events, and enrichment in Fe and light $r$-elements by
L events. So long as different clots of baryonic matter within
the Galaxy underwent the same evolution, this sequence of events can be
established independent of which stars observed today represent the same
initial clot. However, this same sequence of events might have started at
widely different times within different clots. In this sense,
until chemical enrichment became essentially uniform on the Galactic scale,
stellar abundances at low metallicities would only reflect relative time 
in the above sequence. Finally, if galaxies formed $\sim 10^9$~yr after 
the Big Bang, the epoch of supermassive star formation discussed here
would correspond to a redshift of at least $z\sim 4$ [for which
the age of the universe is $\sim 10^{10}(1+z)^{-3/2}\sim 10^9$~yr].
This epoch is earlier than the one probed by recent abundance
observations at ``high'' redshifts ($z\sim 3$, Pettini et al. 1997).

\acknowledgments

We thank Roger Blandford, Wallace Sargent, 
Charles Steidel, and Stan Woosley for helpful discussions
and Roberto Gallino for seeking to keep us on the rich road of Ba
and proper mixing. Comments by Al Cameron and John Cowan 
were very valuable.
This work was supported in part by the Department of Energy under
grants DE-FG03-88ER-13851 and DE-FG02-87ER40328,
and by NASA under grant
NAG 5-4076, Caltech Division Contribution No. 8647(1039).

\clearpage

\figcaption{Data (stars: Gratton \& Sneden 1994; squares: 
McWilliam et al. 1995; McWilliam 1998; triangles: Sneden et al. 1996; 1998)
on [Ba/Eu]$_r\equiv\log({\rm Ba/Eu})
-\log({\rm Ba/Eu})_{\odot,r}$ and $\log\epsilon({\rm Eu})$ versus [Fe/H] 
for low-metallicity stars. The Ba/Eu ratio is near the solar
$r$-process value over the range of [Fe/H] shown. But there is a wide
dispersion in $\log\epsilon({\rm Eu})$ at 
$-3.1\lesssim[{\rm Fe/H}]\lesssim -2.5$. This dispersion disappears for
[Fe/H]~$\gtrsim -2.5$.}

\figcaption{Data (symbols as in Fig. 1)
on $\log\epsilon({\rm Ba})$ versus [Fe/H] for low-metallicity stars.
The range of [Fe/H] extends about one dex below that for
the existing data on Eu. There is a wide dispersion in
$\log\epsilon({\rm Ba})$ at $-3.1\lesssim[{\rm Fe/H}]\lesssim -2.5$.
Three regions of abundance evolution are schematically shown:
production by the initial/prompt Fe source (A), addition of high frequency
non-Fe-producing H events (B), and mixture of H and low frequency
Fe-producing L events (C).}

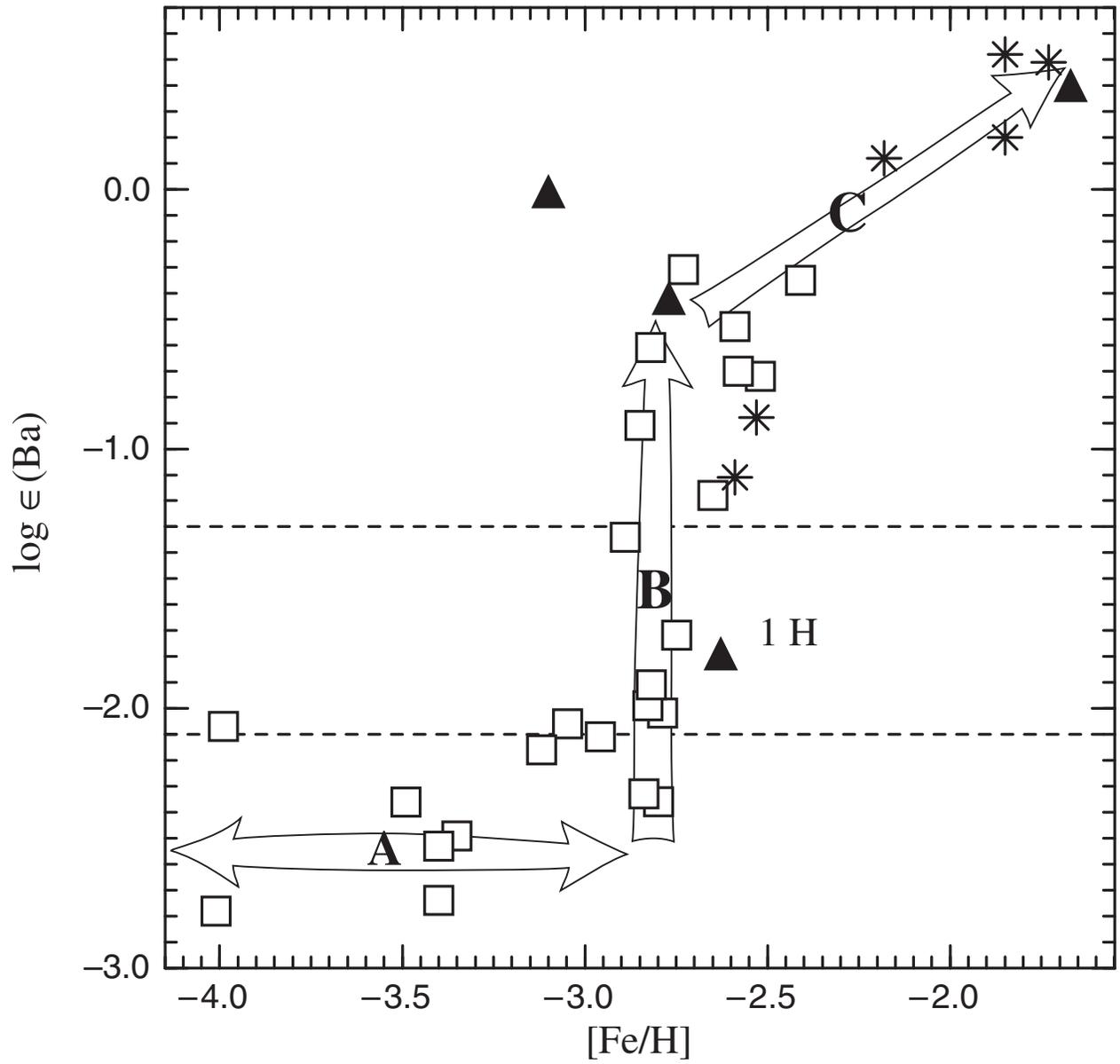
\end{document}